\documentclass[twocolumn, showpacs, preprintnumbers, amsmath, amssymb,noshowpacs, nopreprintnumbers]{revtex4}
\usepackage{pifont}
\usepackage{tikz}
\usetikzlibrary{shapes,arrows}
\usepackage[colorlinks,urlcolor=blue]{hyperref}
\usepackage{epstopdf}

\usepackage{graphicx}% Include figure files
\usepackage{dcolumn}% Align table columns on decimal point
\usepackage{appendix}
\usepackage{breakurl}%
\usepackage{changes}

\makeatletter
  \let\@font@info\@gobble
  \let\@font@warning\@gobble
\makeatother

\begin{document}

%\preprint{APS/123-QED}

\title{Simulation of the continuous-time random walk using subordination schemes}

\author{Danhua Jiang}
\author{Yuanze Hong}

 %\altaffiliation[Also at ]{School of Mathematics and Statistics, Lanzhou University.}%Lines break automatically or can be forced with \\

\author{Wanli Wang}%
\email[contact author:~]{ wanliiwang@163.com}
\affiliation{%
%  $^1$School of Mathematics and Statistics, Gansu Key Laboratory of Applied Mathematics and Complex
%Systems, Lanzhou University, Lanzhou 730000,  P.R. China
%\\
School of Mathematical Sciences, Zhejiang University of Technology, Hangzhou 310023, China
 }%
% \homepage{http://www. Second. institution. edu/~Charlie. Author}
%\affiliation{

%\date{\today}% It is always \today,  today,
\begin{abstract}
The continuous time random walk model has been widely applied in various fields, including physics, biology, chemistry, finance, social phenomena, etc.  In this work, we present an algorithm that utilizes a subordinate formula to generate data of the continuous time random walk in the long time limit.
The algorithm has been validated using commonly employed observables, such as typical fluctuations of the positional distribution, rare fluctuations, the mean and the variance of the position, and breakthrough curves with time-dependent bias, demonstrating a perfect match.
\end{abstract}

%\pacs{02. 50. -r,  05. 20. -y,  05. 40. -a }% PACS,  the Physics and Astronomy
% url
%https://publishing.aip.org/publishing/pacs/pacs-reg00#05
%
%   02. 50. -r   Probability theory, stochastic processes, and statistics
%  05.30.Pr	Fractional statistics systems (anyons, etc.)

% 05.40.-a Fluctuation phenomena, random processes, noise, and Brownian motion (for fluctuations in superconductivity, see 74.40.-n; for statistical theory and fluctuations in nuclear reactions, see 24.60.-k; for fluctuations in plasma, see 52.25.Gj; for nonlinear dynamics and chaos, see 05.45.-a)
%
%??05.10.Gg Stochastic analysis methods (Fokker-Planck, Langevin, etc.)

% 02.50.Ey	Stochastic processes
%  45.10.Hj	Perturbation and fractional calculus methods

\maketitle
\section{Introduction}\label{18ctrwsect1}
The continuous-time random walk (CTRW) \cite{Shlesinger1974Asymptotic,Montroll1965Random,Bouchaud1990Anomalous,Metzler2000random,Gradenigo2016Field,Kutner2017continuous,Liu2022Strong} is a stochastic process for a random walk that jumps from one position to another one.
As an extension of the discrete random walk process, it was originally discussed by
Montroll and Weiss \cite{Montroll1965Random}, where a random walk is subordinated to a renewal process.  Waiting times and displacements of the random particles are independent and identically distributed (IID) \cite{Metzler2000random}. For the waiting times, it may be drawn from a heavy or narrow-tailed distribution. Different combinations of waiting times and displacements distributions yield different types of diffusion statistics, ranging from sub-diffusion, normal diffusion, and super diffusion. The CTRW model has attracted great attention from researchers due to its wide applications.  For example, it was used in many physical and chemical phenomena, including the transport of amorphous materials \cite{Scher1975Anomalous}, contamination in disorder systems \cite{Alon2017Time}, the U.S. dollar in Deutsche mark future exchange \cite{Masoliver2003Continuous}, diffusion of polymers in attractive nano-particle polymer mixtures \cite{Hu2023Triggering} and so on.

The key issue is to discuss and simulate the positional distribution of the CTRW model. Two alternative approaches are the  fractional diffusion equation \cite{Metzler2000random,Rehim2008Simulation} and Langevin equations \cite{Magdziarz2009Langevin,Akimoto2016Distributional,Majumdar2015Effective}. For example, Fogedby used decouple Langevin equations concerning power law step and waiting time distributions \cite{Fogedby1994Langevin} to illustrate the CTRW model. One advantage is that the drift force field of the system is natural to be added. This method was further extended in Ref.~\cite{Kleinhans2007Continuous,Richmond2019Simulation}. See also discussions of coupled cases \cite{Magdziarz2015Asymptotic}. Here we use the statistics of the number of renewals in the long time limit, eliminating the need to generate waiting times statistics.

This manuscript focuses on the case when waiting time has a finite mean and an infinite variance, and the displacement follows a Gaussian distribution with a non-zero mean. However, the methods we provided are much more general. Due to the asymmetric displacement distribution, sometimes it was called the biased CTRW model. The biased CTRW has garnered significant attention and a vast number of specialized literatures \cite{Bel2005Occupation,Burioni2014Scaling,Hou2018Biased,Akimoto2018Ergodicity,Michelitsch2020Biased,Burov2022Exponential}.
When the bias of the system or the mean of the displacement is zero, the process shows normal diffusion. While, when the bias is added, CTRW shows the enhanced or supper diffusion \cite{Shlesinger1974Asymptotic}, i.e.,
\begin{equation*}
\langle x^2\rangle-\langle x\rangle^2  \sim t^{3-\alpha}
\end{equation*}
with $1<\alpha<2$.
In the long time limit, the positional distribution follows the asymmetric L{\'e}vy stable law \cite{Kotulski1995Asymptotic} according to the Montroll-Weiss equation \cite{Bouchaud1990Anomalous,Metzler2000random}. The fat tail is a key characteristic of the asymmetric L{\'e}vy distribution.   Generally, when seeking data on CTRW, we obtain a random count representing the number of renewals occurring within the time interval from zero to $t$, then use subordinated processes to get the  particle's position. A drawback is that it takes a long time to find the number of renewals when the observation time is long.  Thus, it would be interesting to find a new way to generate the number of renewals and the position of particles.

Recall that in the long time limit, the number of renewals follows the asymmetric L{\'e}vy stable law \cite{Godreche2001Statistics}. By instinct, one possible way is to generate the number of renewals based on the mentioned L{\'e}vy stable law. The L{\'e}vy stable law is perfectly correct, but it does have some drawbacks. For example, the far tail of L{\'e}vy distribution tends to infinity, which indicates that the number of the renewals tends toward negative infinity.  In turn, the L{\'e}vy stable law gives an infinite mean square displacement \cite{Godreche2001Statistics,Wang2018Renewal} (MSD) of the number of renewals. It is certainly not true, as the far tail should have a cutoff \cite{Wang2018Renewal} for a finite observation time $t$.  In addition, as suggested in \cite{Wang2018Renewal}, the MSD of the number is governed by an infinite density. On the other hand, the positional distribution in the context of simulations converges to the L{\'e}vy stable law rather slowly, especially when the system's bias is weak.
In some sense, the convergence problem was solved in \cite{Wang2020Fractional} using the fractional advection-diffusion-asymmetry equation. To be more exact, the solution of the equation is the convolution of the L{\'e}vy stable law and a Gaussian distribution. Unfortunately, the solution of the fractional advection-diffusion-asymmetry equation yields a divergent MSD \cite{Wang2020Fractional,Wang2024Fractional}.

Our goal in this manuscript is to construct a PDF to describe the number of renewals, which is not only valid for the typical fluctuations of the number of renewals but also effective for rare events. The main idea is that we use the L{\'e}vy stable law  and the non-normalized density to illustrate the central part and the far tail of the number of renewals, respectively. Below, we will demonstrate that these two laws exhibit distinct scaling limits on the number of renewals and then utilize commonly employed observables to illustrate the effectiveness of the constructed PDF.

The organization of the manuscript is as follows.  In Sec. \ref{set2}, we give the definition of the CTRW model and the algorithm to generate the number of renewals in the long time limit. Specifically, the modified L{\'e}vy stable law is introduced. We further check the theoretical prediction using the widely used observables in Sec. \ref{Sim3}, including  typical fluctuations of the position, large deviations of the position describing the far tail, the mean and the variance of the position, and breakthrough curves with the time-dependent bias. Finally, we conclude the manuscript with a discussion in Sec. \ref{24seccon}.

\section{Model and algorithm}\label{set2}
Below, we present the definition of the CTRW model and the algorithm for simulating data on the number of renewals and positions.

\subsection{Model}

In this subsection, we introduce the CTRW model, which is defined as follows. A random walk starts from its initial position. The particle waits at its initial position for time $\tau_1$ drawn from $\phi(\tau)$, and then makes a jump to $x_1$ with $x_1$ following $f(\chi)$. The particle traps at $x_1$ for time $\tau_2$ and then jumps to the position $x_1+x_2$ with $x_2$ generated from $f(\chi)$. Then, the process is renewed.  All along the manuscript, the waiting times $\tau_i$ are IID random variables, and the same holds for $\chi_i$. The number of renewals, denoted as $N_t$, obeys the relation $\sum_{i=1}^{N_t}\tau_i+B_t=t$, where $B_t$ is the backward recurrence \cite{Wang2018Renewal,Godreche2001Statistics} and $t$ is the total observation time.

All along the manuscript, waiting times follow a power law distribution \cite{Wang2018Renewal,Klafter2011First,Deng2020Modeling}
\begin{equation}\label{23eqs100}
\displaystyle \phi(\tau)=\left\{
          \begin{array}{ll}
            0, & \hbox{$\tau<\tau_0$;} \\
            \alpha\frac{\tau_0^\alpha}{\tau^{1+\alpha}}, & \hbox{$\tau\geq\tau_0$}
          \end{array}
        \right.
\end{equation}
with $1<\alpha<2$. As mentioned, waiting times  $\tau$ have a finite mean and an infinite variance. Performing the Laplace transform with respect to $\phi(\tau)$, we have $\widehat{\phi}(s)=\int_0^\infty \exp(-st)\phi(\tau)d\tau\sim 1-\langle\tau\rangle s+(\tau_0)^\alpha|\Gamma(1-\alpha)|s^\alpha$ with $\langle\tau\rangle=\alpha\tau_0/(\alpha-1)$ being the mean of waiting times.

For displacements, we focus on Gaussian distribution with a finite mean $a$ and a finite variance $\sigma^2$
\begin{equation}\label{23eqs101}
  f(x)=\frac{1}{\sqrt{2\sigma^2\pi}}\exp\left[-\frac{(x-a)^2}{2\sigma^2}\right].
\end{equation}

Let $Q_t(N)$ be the PDF of the number of renewals and $f(x|N)$ be the conditional probability of finding the particle on the position $x$ conditioned it made $N$ jumps. The PDF of the position follows
\begin{equation}\label{suborb}
\begin{split}
P(x,t)&=\sum_{N=0}^{\infty}Q_t(N)f(x|N)\\
&\rightarrow \int_{0}^{\infty}Q_t(N)f(x|N){\rm d}N.
\end{split}
\end{equation}
Eq.~\eqref{suborb} describes the particle at position $x$ after $N$ steps on conditional that  exactly $N$ steps were
made up to time $t$.  Eq.~\eqref{suborb} can be interpreted as the so-called integral formula
of subordination \cite{Bouchaud1990Anomalous,Klafter1994Probability,Barkai2001Fractional,Wang2020Fractional}, i.e., a random walk is subordinated to a renewal process.
The subordination scheme has been employed in mathematics and physics within various contexts, including fractional kinetic equations \cite{Alexander1997Fractional,Klafter2011First,Chechkin2021Relation}, Jacobi stochastic volatility models \cite{tong2022pricing}, and financial models \cite{semeraro2022multivariate}. Below, we will use Eq.~\eqref{suborb} to generate the data of positions.

\begin{figure}[htb]
 \centering
 \includegraphics[width=0.5\textwidth]{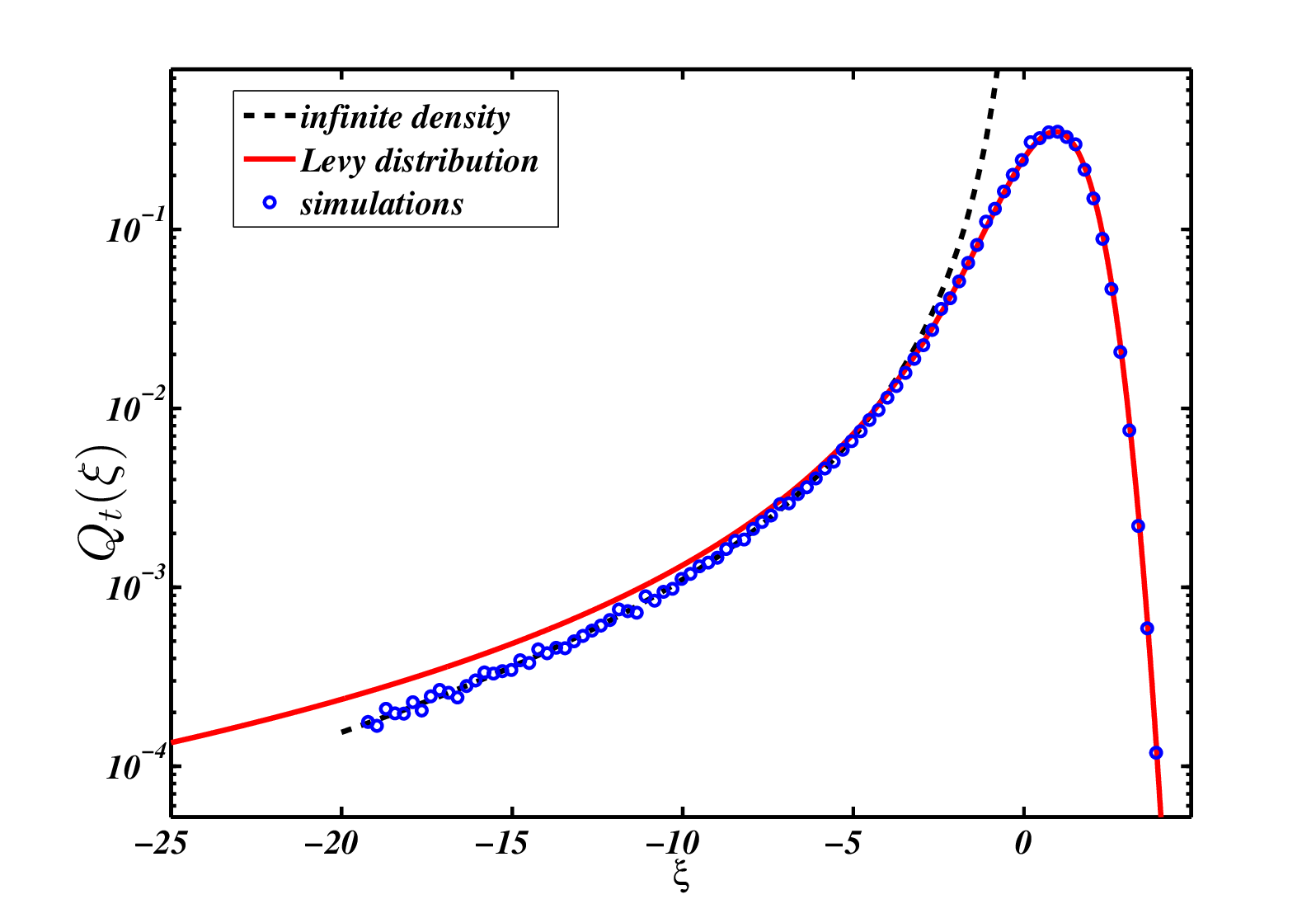}\\
 \caption{Distribution of the number of renewals. The red solid line describes typical fluctuations of the number of renewals, shown by the L{\'e}vy law Eq.~\eqref{24sec101} and rare events Eq.~\eqref{app102} are plotted by the black dashed line. Symbols plotted by circles are simulations of the renewal process for $2\times 10^7$ realizations. We choose $t=1000, \tau_0=0.1$, and $\alpha=3/2$.
}\label{LevyLaw}
\end{figure}

 Recall that  displacements are IID random variables, thus $f(x|N)$ obeys
\begin{equation}\label{23eqs403}
f(x|N)=\frac{1}{\sqrt{2\sigma^2\pi N}}\exp\left[-\frac{(x-aN)^2}{2\sigma^2N}\right].
\end{equation}
As described in \cite{Kotulski1995Asymptotic,Godreche2001Statistics,Wang2018Renewal}, in the long time limit, the number of renewals follows the asymmetric L{\'e}vy distribution
\begin{equation}\label{24sec101}
Q_t(\xi)\sim \mathcal{L}_{\alpha}(\xi)
\end{equation}
with $\xi=(N-t/\langle\tau\rangle)/(t/\bar{t})^{1/\alpha}$  and $\bar{t}=\langle\tau\rangle^{1+\alpha}/(\tau_0^\alpha|\Gamma(1-\alpha)|)$. Eq.~\eqref{24sec101} is plotted by the red solid line in Fig.~\ref{LevyLaw}.
Here the symbol $\mathcal{L}_{\alpha}(\xi)$ is the L{\'e}vy stable distribution, whose Fourier transform is $\int_{-\infty}^{\infty}\exp(-ik\xi)\mathcal{L}_{\alpha}(\xi)dx=\exp((-ik)^\alpha)$. Typical fluctuations Eq.~\eqref{24sec101} describes the case when $N-t/\langle\tau\rangle$ is of the order of $t^{1/\alpha}$, i.e., the central part of the distribution. Using the asymptotic behavior of the L{\'e}vy distribution, the far  left tail of $N$ follows $Q_t(\xi)\sim (-\xi)^{-1-\alpha}/\Gamma(-\alpha)$. It indicates that Eq.~\eqref{24sec101} does not give the information of  the MSD \cite{Wang2018Renewal}.

\subsection{Algorithm}\label{Algo}

According to  Eqs.~(\ref{suborb}, \ref{23eqs403}, \ref{24sec101}), in the long time limit, the positional distribution follows
\begin{equation}\label{18eq502}
\begin{split}
P(x,t)\sim& \frac{1}{(t/\overline{t})^{1/\alpha}}\int_{0}^{\infty}\mathcal{L}_{\alpha}\left(\frac{N-t/\langle\tau\rangle}{(t/\overline{t})^{1/\alpha}}\right)\\
    & \frac{\exp\Big(-\frac{(x-aN)^2}{2\sigma^2 N}\Big)}{\sqrt[]{2\pi\sigma^2 N}}{\rm d}N,
\end{split}
\end{equation}
Changing the random variables, the above equation leads to
\begin{equation}\label{18eq503}
\begin{split}
P(x,t)\sim &\int_{-\frac{t}{\langle\tau\rangle}(\overline{t}/t)^{1/\alpha}}^{\infty}\mathcal{L}_{\alpha}(\xi)\\
    & \times \frac{\exp\Big(-\frac{(x-a\frac{t}{\langle\tau\rangle}-a\xi(\frac{t}{\overline{t}})^{1/\alpha})^2}{2\sigma^2(t/\langle\tau\rangle+\xi(t/\overline{t})^{1/\alpha})}\Big)}{\sqrt[]{2\sigma^2\pi(\frac{t}{\langle\tau\rangle}+\xi(\frac{t}{\overline{t}})^{1/\alpha})}}{\rm d}\xi,
\end{split}
\end{equation}
Eq.~\eqref{18eq503} is the main result of this section used to generate the random variables. Note that $\mathcal{L}_{\alpha}(\xi)$ in Eq.~\eqref{18eq503} should be modified, denoted as $\mathcal{L}_{\alpha}^{*}(\xi)$.
Otherwise, it will not accurately account for the MSD and the far tail of the position distribution.
For that, we generate the statistics of $\mathcal{L}_{\alpha}^{*}(\xi)$, stemming from typical fluctuations and the infinite density \cite{Erez2017Large} and then use it to get the data of the position. Afterwards, positions of particles are drawn from a Gaussian distribution with the mean $at/\langle\tau\rangle+a\xi(t/\bar{t})^{1/\alpha}$ and the variance $\delta^2(t/\langle\tau\rangle+\xi(t/\bar{t})^{1/\alpha})^2$.

Rare fluctuations of $N$ were investigated, describing the scaling when $N-t/\langle\tau\rangle$ is of the order of $t$. According to the result given in \cite{Wang2018Renewal}, large deviations of $N$ obey
\begin{equation}\label{app101}
Q_t(\epsilon)\sim \alpha(\tau_0)^\alpha\left[t(-\epsilon\langle\tau\rangle)^{-\alpha-1}+\frac{1-\alpha}{\alpha}(-\epsilon\langle\tau\rangle)^{-\alpha}\right]
\end{equation}
with $\epsilon=N-t/\langle\tau\rangle$.  Consider the random variable, $\xi=\epsilon/(t/\bar{t})^{1/\alpha}$, we have
\begin{equation}\label{app102}
\begin{split}
  Q_t(\xi)\sim & \alpha(\tau_0)^\alpha\left(\frac{t}{\bar{t}}\right)^{1/\alpha}\\
    & \times \left[\frac{t}{(-\xi(\frac{t}{\bar{t}})^{1/\alpha}\langle\tau\rangle)^{\alpha+1}}+\frac{(1-\alpha)/\alpha}{(-\xi(\frac{t}{\bar{t}})^{1/\alpha}\langle\tau\rangle)^{\alpha}}\right].
\end{split}
\end{equation}
See the black dashed line in Fig.~\ref{LevyLaw}.
Recall that Eq.~\eqref{app102} is valid for small $\xi$ corresponding to small $N$.
As shown in Fig. \ref{LevyLaw}, Eq.~\eqref{24sec101} is valid for the central part of the number of renewals together with the right tail, and the non-normalized density Eq.~\eqref{app102} works for the left far tail. By instinct, we can utilize the effective part of Eqs.~\eqref{24sec101} and \eqref{app102} to generate data of the number of renewals.

In order to generate the random variable based on Eq.~\eqref{app102}, we construct a PDF by adding a condition $z$ on Eq.~\eqref{app102},
\begin{equation}\label{app103}
Q_t^{I}(\xi)=\alpha(\tau_0)^\alpha\left[\frac{\bar{t}}{(-\xi\langle\tau\rangle)^{\alpha+1}}+\frac{(1-\alpha)/\alpha}{(-\xi\langle\tau\rangle)^{\alpha}(\frac{t}{\bar{t}})^{1-\frac{1}{\alpha}}}\right]
\end{equation}
with $\bar{b}<\xi\leq z<0$. Here, $z$ is an undetermined constant.
As $N\geq 0$, the bottom limit of $\xi$ should be great than $-t^{1-1/\alpha}\bar{t}^{1/\alpha}/\langle\tau\rangle$, i.e.,
\begin{equation}
\bar{b}=-\left(\frac{\alpha}{\Gamma (2-\alpha)}\right)^{1/\alpha} \left(\frac{t}{\tau_0}\right)^{1-\frac{1}{\alpha}}.
\end{equation}
When $1<\alpha<2$, in the long time limit, $\bar{b}\to -\infty$. While for a finite observation time $t$, $\xi$ can not tend to infinity. Thus, the modified L{\'e}vy stable law $\mathcal{L}_\alpha^{*}(\xi)$ should have a cutoff at the tail.

We further introduce the cumulative distribution function, describing the probability of $\xi$ when $\xi $ is less than $y$
\begin{equation}\label{app104}
h(y)=\int_{\bar{b}}^y Q_t^{I}(\xi){\rm d}\xi.
\end{equation}
Performing the integral from $\bar{b}$ to $z$, Eq.~\eqref{app104} leads to
\begin{equation}\label{app105}
\frac{\tau_0 z ((\alpha-1) t)^{1/\alpha} | \Gamma (1-\alpha)| ^{1/\alpha}+t (\alpha \tau_0)^{1/\alpha}}{t (\alpha \tau_0)^{1/\alpha} (-z)^{\alpha} | \Gamma (1-\alpha)| }=1.
\end{equation}
Eq.~\eqref{app105} can be used to determine the upper limit
$z$ defined in Eq.~\eqref{app103}. When $\alpha=3/2$, using a cubic equation, the formula
for the roots of Eq.~\eqref{app105} is easy to find. While, if $\alpha=5/4$, Eq.~\eqref{app105} reduces to a $5$-th degree polynomial. Thus, according to the Abel-Ruffini theorem, the equation concerning $z$ shown by Eq.~\eqref{app105} is generally unsolvable. For that, to have a universal program valid for all kinds of $\alpha$, we have to abandon some methods that are $z-$ dependent.

\begin{figure}[htb]
 \centering
 \includegraphics[width=0.5\textwidth]{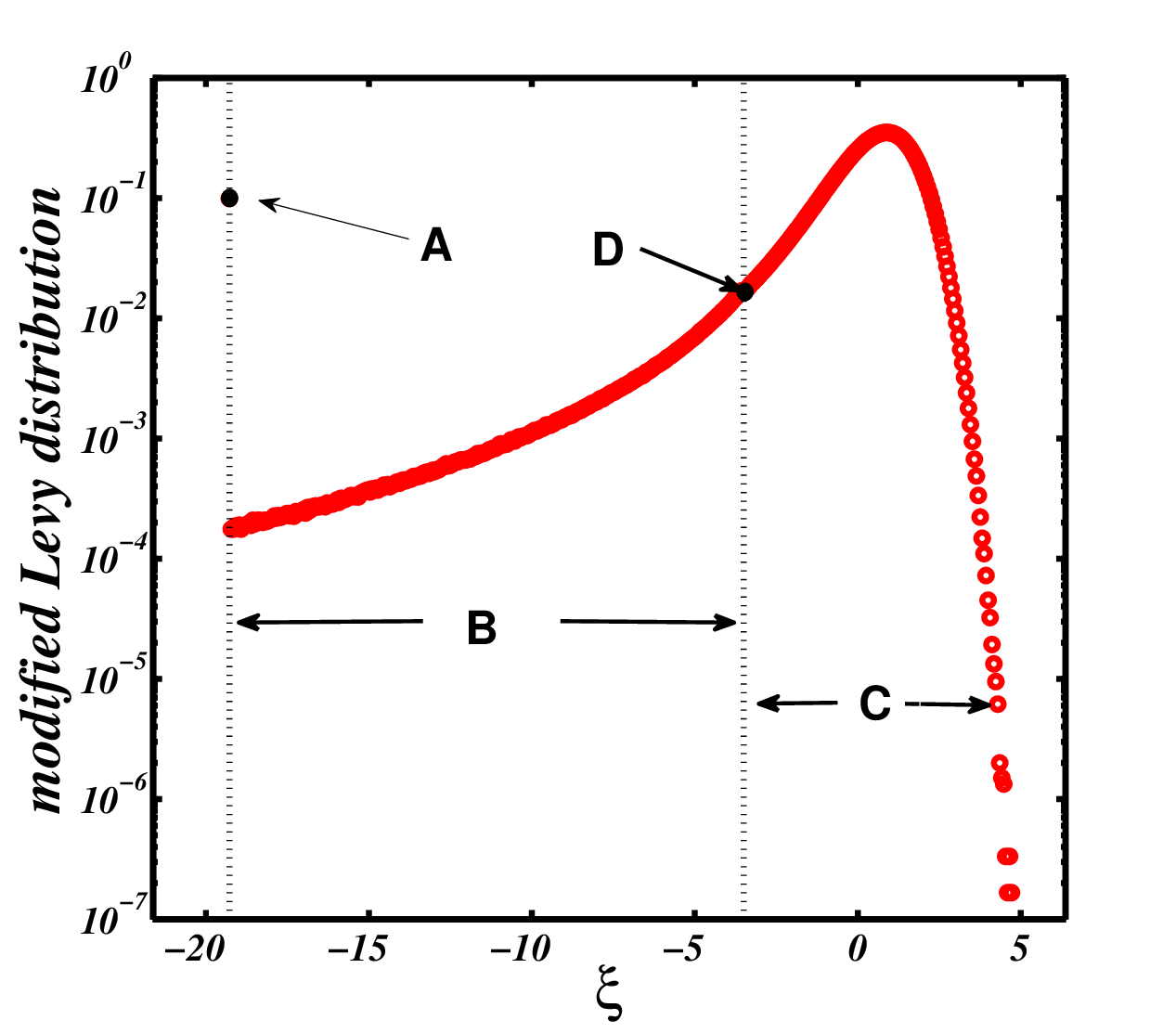}\\
 \caption{Plot of the modified asymmetric L{\'e}vy stable distribution. Here, the black dot marked by A describes the modified survival probability Eq.~\eqref{app108}, the part B stems from the modified infinite density Eq.~\eqref{app103}, the part C is related to typical fluctuations Eq.~\eqref{24sec101}, and the black dot D denoted as $\xi^{*}$ is  the crossover point between typical fluctuations and rare fluctuations.
}\label{modifiedLevy}
\end{figure}

As shown in Fig.~\ref{modifiedLevy}, the modified L{\'e}vy stable has three parts, labeled as $A$, $B$, and $C$, respectively. Part $A$ represents the  probability for the normalization, describing the fade probability of random variables equaling to $\bar{b}$. For part $B$, we utilize the information of the modified infinite density Eq.~\eqref{app103}.
In part $C$, random variables $\xi$ are  drawn from the L{\'e}vy stable law Eq.~\eqref{24sec101}.
The dot $D$ is the intersection of the part $B$ and the part $C$.

Below, we calculate the probability of random variables located in regions B and C, respectively.  For the probability of rare fluctuations in region B, we have
\begin{equation}\label{app106}
\begin{split}
P_{\rm{rare}}(\xi^{*})&=h(\xi^{*})\\
&=\frac{\tau_0 z ((\alpha-1) t| \Gamma (1-\alpha)|)^{1/\alpha}+t (\alpha \tau_0)^{1/\alpha}}{t (\alpha \tau_0)^{1/\alpha} (-\xi^{*})^{\alpha} | \Gamma (1-\alpha)| },
\end{split}
\end{equation}
as mentioned before, $\xi^{*}$ is the transition point of the region B and the region C. Recall that part C is illustrated by the L{\'e}vy stable law, the probability in the region C follows
\begin{equation}\label{app107}
P_{\rm{typical}}(\xi^{*})=\int_{\xi^{*}}^\infty \mathcal{L}(\xi){\rm d}\xi.
\end{equation}
When $t\to\infty$, $P_{\rm{rare}}(\xi^{*})+P_{\rm{typical}}(\xi^{*})$ tends to unity. While for a finite $t$, $P_{\rm{rare}}(\xi^{*})+P_{\rm{typical}}(\xi^{*})<1$. Therefore, we further introduce the modified or fake survival probability describing the non-traveling particles
\begin{equation}\label{app108}
P_{\rm{survival}}(\xi^{*})=1-P_{\rm{rare}}(\xi^{*})-P_{\rm{typical}}(\xi^{*}).
\end{equation}
In other words, it is related to scenarios where the number of renewals is zero, which leads to the non-moving particles in the context of the CTRW model.

Now we discuss the details of generating random variables $\xi$ drawn from $\mathcal{L}_{\alpha}^{*}(\xi)$. Firstly, we generate a random variable drawn from a uniform distribution, denoted as $\eta$. Based on the value of
$\eta$, we determine which part it belongs to.
If $\eta > P_{\rm{survival}}(\xi^{*}) + P_{\rm{rare}}(\xi^{*})$, we generate a random variable $\xi$ from $\mathcal{L}(\xi)$ under the condition that $\xi \geq \xi^{*}$. Instructions for generating a L{\'e}vy stable distribution can be found on \cite{Chambers1976Methods,Janicki1995Computer}
or the MATLAB Central File Exchange, as detailed in the work by Veillette \cite{veillette2012stbl}.
When $P_{\rm{survival}}(\xi^{*}) < \eta < P_{\rm{survival}}(\xi^{*}) + P_{\rm{rare}}(\xi^{*})$,  $\xi$ are drawn from the modified infinity density Eq.~\eqref{app103} using the accept/reject algorithm. If $\eta < P_{\rm{survival}}(\xi)$, the random variable is fixed, i.e., $\xi = \bar{b}$. Note that the MSD is sensitive to lagging particles far behind the mean \cite{Wang2019Transport}. Therefore, when simulating the position for the MSD, the random variables $\eta$ are larger than $P_{\rm{survival}}(\xi^{*})$ using the theory given in Eq.~\eqref{app103}. Throughout the manuscript, the theories marked in  legends of figures refer to the algorithm discussed in this subsection. Specifically, we generate $\xi$ drawn from the modified L{\'e}vy law $\mathcal{L}_{\alpha}^{*}(\xi)$ and then use it to determine the position statistics given by the expression
$$J(x)=\frac{\exp\Big(-\frac{(x-a\frac{t}{\langle\tau\rangle}-a\xi(\frac{t}{\overline{t}})^{1/\alpha})^2}{2\sigma^2(t/\langle\tau\rangle+\xi(t/\overline{t})^{1/\alpha})}\Big)}{\sqrt[]{2\sigma^2\pi(\frac{t}{\langle\tau\rangle}+\xi(\frac{t}{\overline{t}})^{1/\alpha})}}.$$
Here, $J(x)$ represents the distribution of the position
$x$ at the time $t$ within the context of the CTRW model, where  $\xi$ are drawn from $\mathcal{L}_{\alpha}^{*}(\xi)$.

%\begin{figure}[htb]
% \centering
% % Requires \usepackage{graphicx}
% \includegraphics[width=0.50\textwidth]{Fig/GenerateProgress.eps}\\
% \caption{The statistics positional distribution.
%}\label{GenerateProgress}
%\end{figure}

\section{Theoretical predictions compared with asymptotic behaviors and simulations}\label{Sim3}

In this section, we generate the data of the position obtained from the method outlined in Sec.~\ref{Algo} to obtain  typical fluctuations of the position, rare fluctuations, the mean of position, the variance of the position, and breakthrough curves. Additionally, we generate the positional distribution data using simulations and outline the well-known asymptotic behaviors for comparative analysis.

\subsection{Typical fluctuations}
\begin{figure}[htb]
 \centering
 \includegraphics[width=0.5\textwidth]{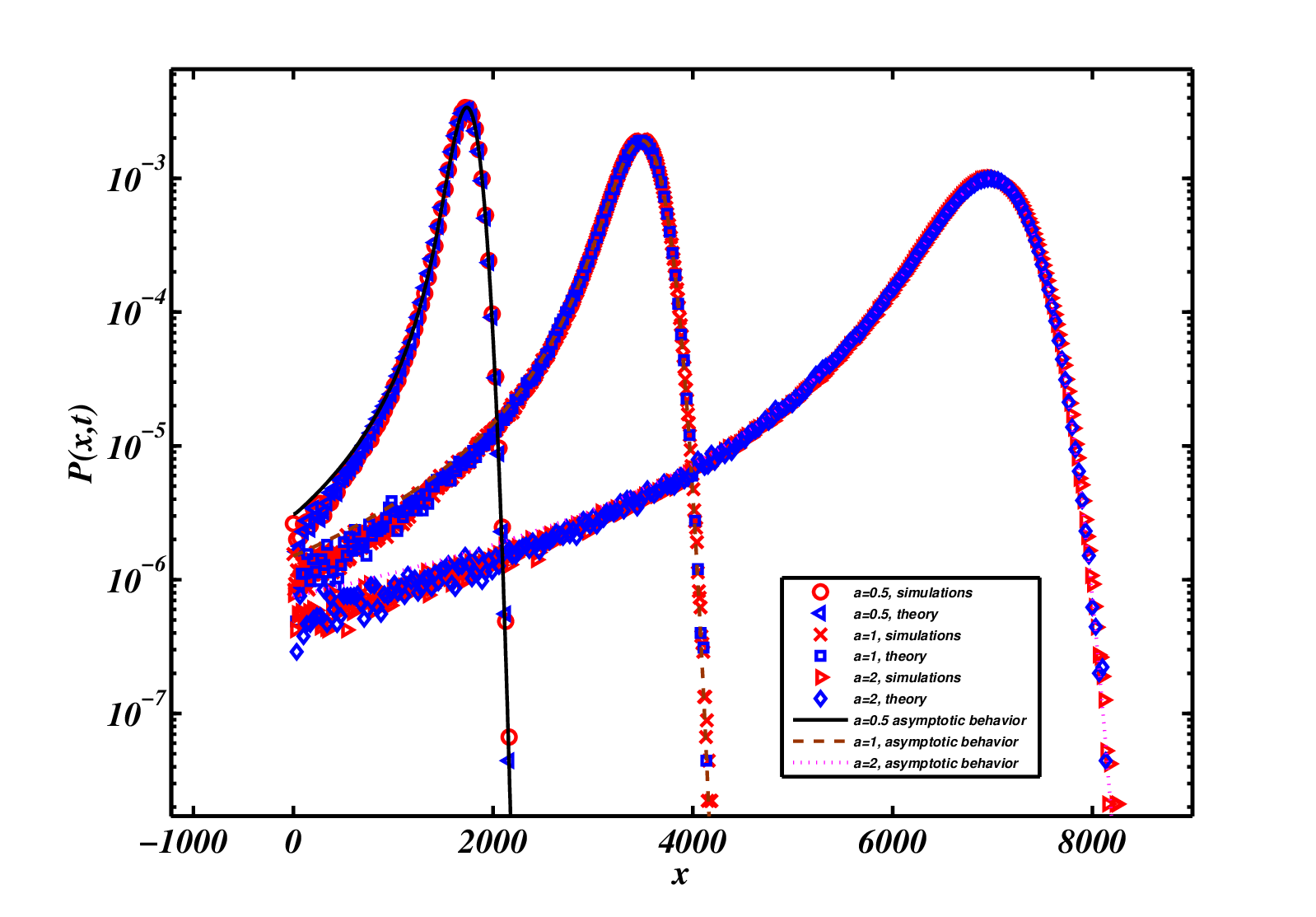}\\
 \caption{Distribution of the position for different biases. Here, the red symbols describe simulations and the green symbols indicate theoretical predictions. For the theories, we use the algorithm provided in Sec.~\ref{Algo}.  The solid lines stem from Eq.~\eqref{23eqs401} describing typical fluctuations of $x$.
  We choose $\alpha=3/2, \sigma=1, t=1000$, and $\tau_0=0.1$.
}\label{PxofLog}
\end{figure}

\begin{figure}[htb]
 \centering
 \includegraphics[width=0.5\textwidth]{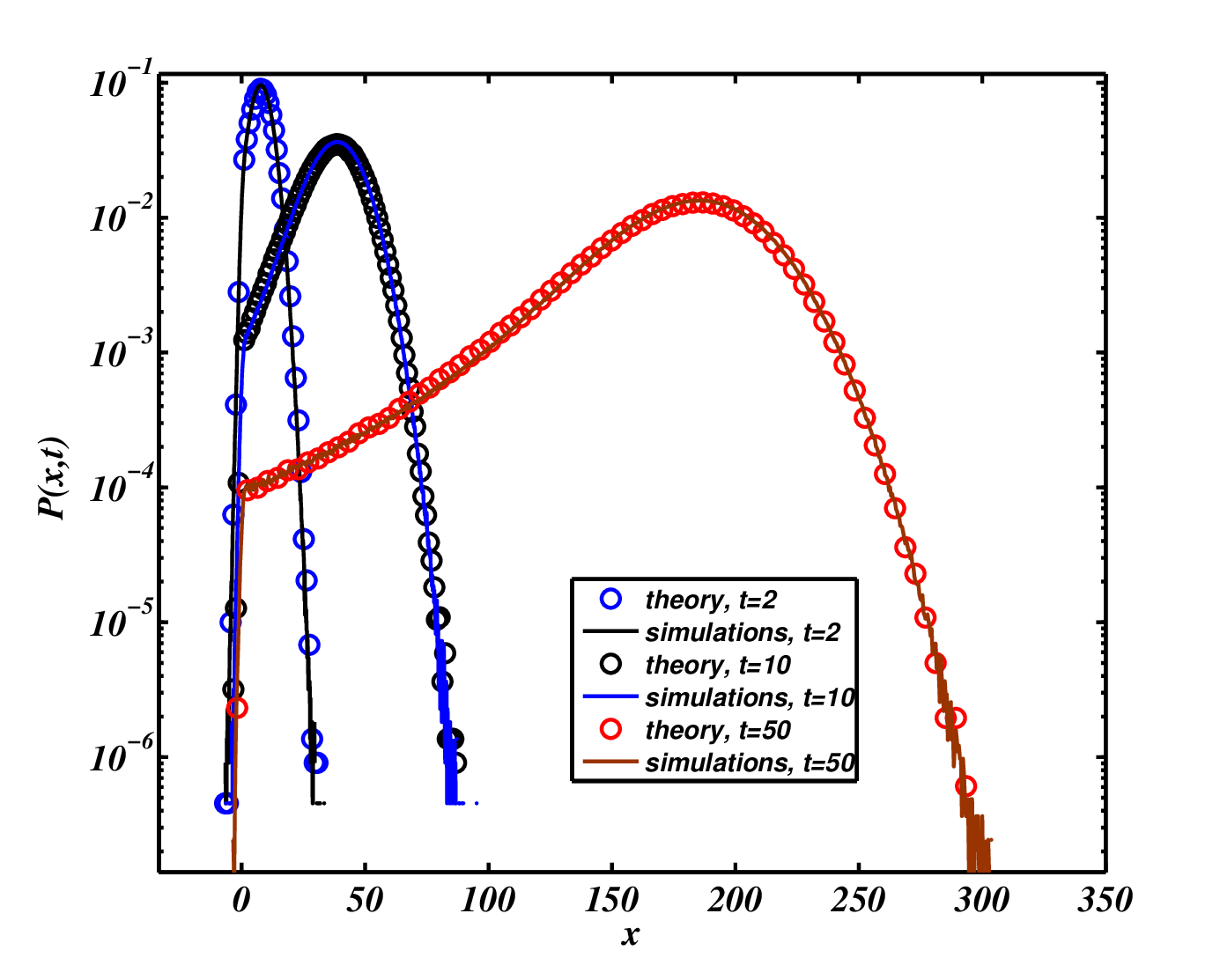}\\
 \caption{Positional distribution with a short observation time $t$, i.e., $t=2, 10$, and $50$. The symbols are theoretical predictions, and the corresponding simulations are plotted using the solid lines. Here we choose $\alpha=3/2, \tau_0=0.1, \delta=1, a=1$, and $2\times 10^7$ realizations.
}\label{GaussianLog}
\end{figure}

\begin{figure}[htb]
 \centering
 \includegraphics[width=0.5\textwidth]{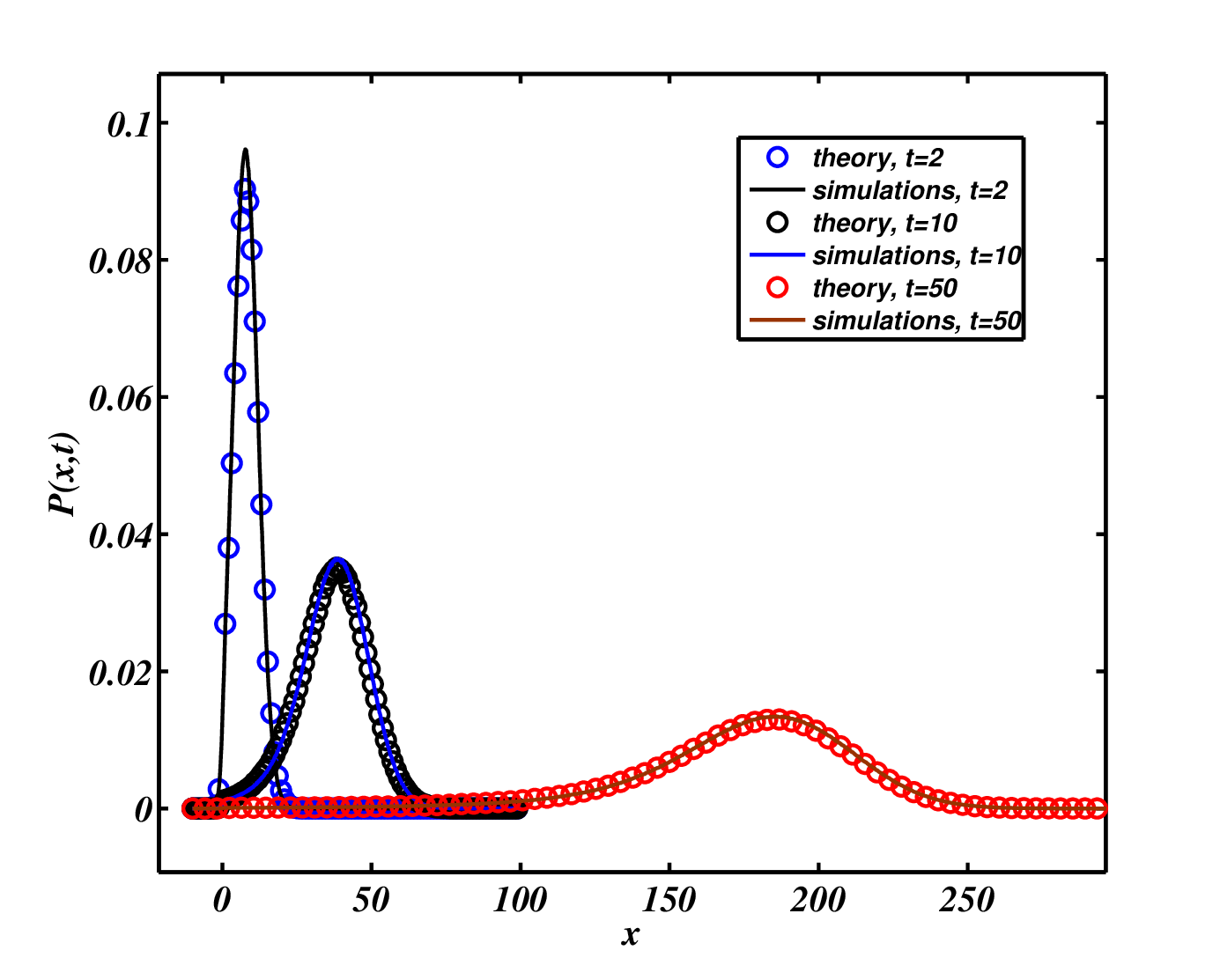}\\
 \caption{Same as in Fig.~\ref{GaussianLog} in linear scale.
}\label{GaussianLinear}
\end{figure}

Several authors have studied typical fluctuations of the position for the biased CTRW; see Refs. \cite{Kotulski1995Asymptotic,Burioni2014Scaling,Wang2019Transport,Alessandro2019Single,Wang2020Fractional}.  In the long-time approximation, the scaled positional distribution follows the asymmetric L{\'e}vy stable distribution \cite{Kotulski1995Asymptotic,Burioni2013Rare}
\begin{equation}\label{23eqs400}
P(\zeta)\sim \mathcal{L}_{\alpha}(\zeta)
\end{equation}
with $\zeta=(x - at/\langle\tau\rangle)/l(t) \rangle$ and $l(t) = a(t/\overline{t})^{1/\alpha}$. In Eq.~\eqref{23eqs400}, it is crucial to emphasize that $a$ must be nonzero; otherwise, the positional distribution would yield a Gaussian distribution. In the general scenario, $P(x,t)$ is described by the convolution of a L{\'e}vy stable distribution and a Gaussian distribution \cite{Wang2020Fractional}
\begin{equation}\label{23eqs401}
P(x,t)\sim \int_{-\infty}^\infty \mathcal{L}_{\alpha}(\xi)\frac{\exp\left(-\frac{(x-\frac{at}{\langle\tau\rangle}-a\xi(\frac{t}{\overline{t}})^{1/\alpha})^2}{2\sigma^2\frac{t}{\langle\tau\rangle}}\right)}{\sqrt{2\pi\sigma^2\frac{t}{\langle\tau\rangle}}}{\rm d}\xi,
\end{equation}
describing typical fluctuations when $x-at/\langle\tau\rangle\sim t^{1/\alpha}$, i.e., the central part of the positional distribution. As shown in Fig.~\ref{PxofLog}, Eq.~\eqref{23eqs401} fails for the left tail of the positional distribution. While the theoretical prediction given in the manuscript is valid for both tails. It indicates that if we use the modified L{\'e}vy stable law in Eq.~\eqref{18eq503}, Eq.~\eqref{18eq503} does work perfectly. In addition, we plot the positional distribution at a short time $t$, see Figs.~\ref{GaussianLog} and \ref{GaussianLinear}. For this case, the positional distribution follows a nearly Gaussian distribution, affirming the validity of our findings. All along the manuscript, the theoretical predictions are based on the algorithm described in Sec.~\ref{Algo}, which uses the modified L{\'e}vy stable law to generate statistics of the number of renewals.

\begin{figure}[htb]
 \centering
 \includegraphics[width=0.5\textwidth]{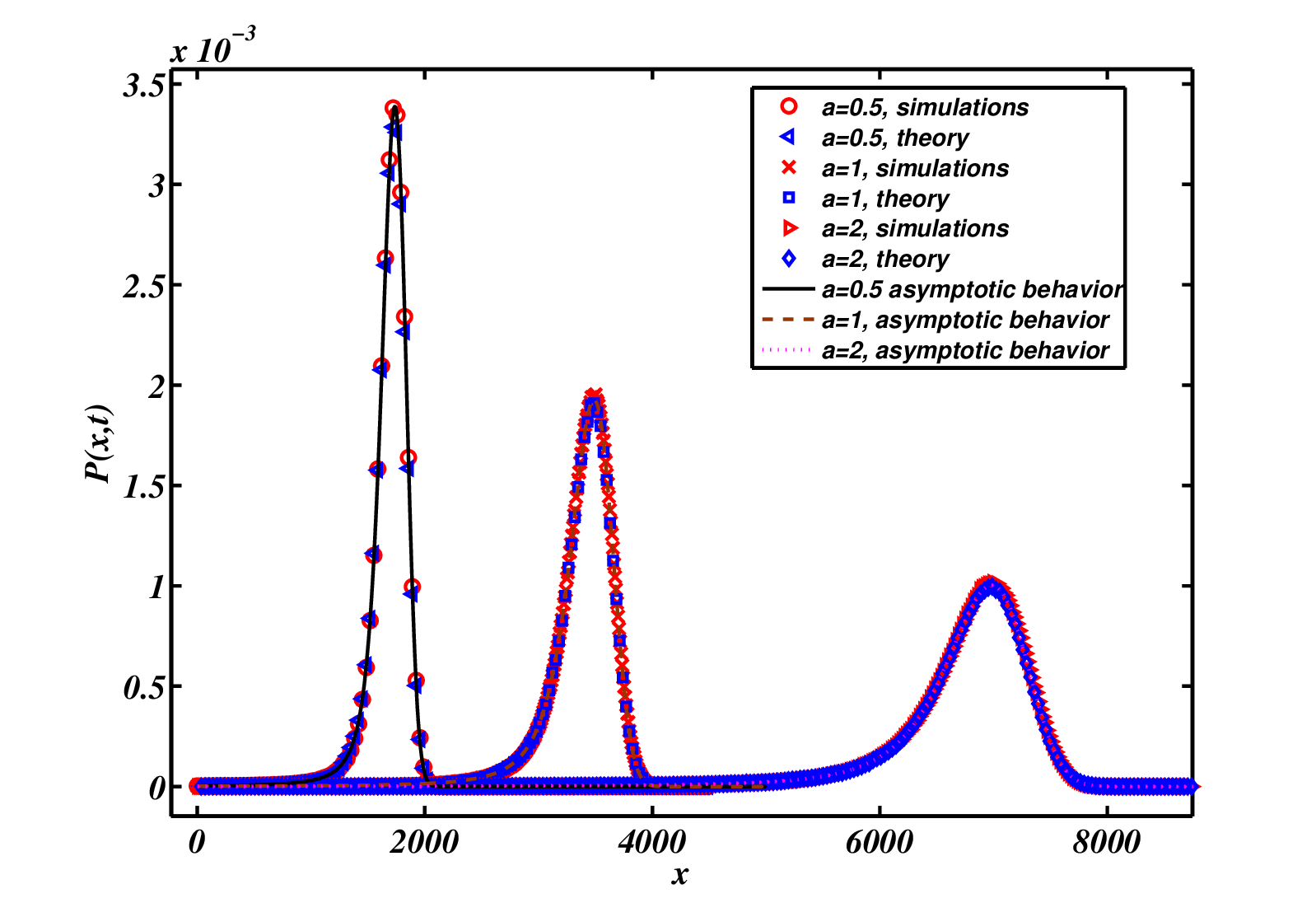}\\
 \caption{Same as in Fig.~\ref{PxofLog} in linear scale.
}\label{PxofLinear}
\end{figure}

\subsection{Rare fluctuations}\label{701}
\begin{figure}[htb]
 \centering
 \includegraphics[width=0.5\textwidth]{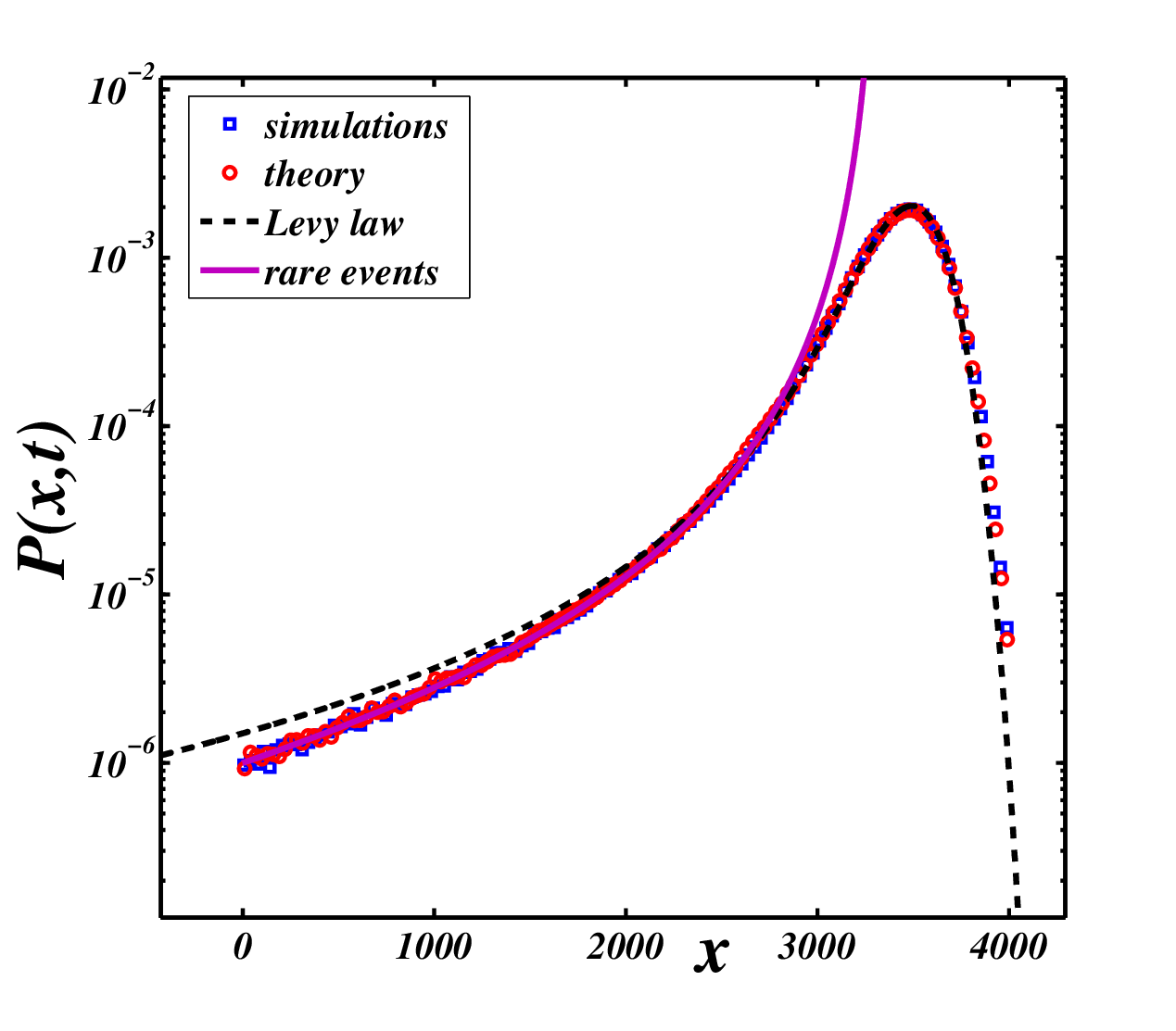}\\
 \caption{Simulations ('$\Box$') of the positional distribution are compared to rare fluctuations Eq.~\eqref{rare01} (solid line), the L{\'e}vy law Eq.~\eqref{23eqs400} in terms of $x$ (dashed line), and theory ('$\circ$').
The parameters are $a=1$, $\sigma=1$, $t=1000$, $\alpha=3/2$, $\tau_0=0.1$, and $2\times10^7$ realizations.
}\label{RareEvents}
\end{figure}

To proceed, we discuss the limit of the Montroll-Weiss equation \eqref{23eqs1000}  but when $|s|/|k|$ is fixed. Specifically, $s$ and $k$ are small and in the same order.  As a  complementary to the L{\'e}vy law Eq.~\eqref{23eqs400}, the rare fluctuations illustrate the far tail of the positional distribution.
Following the result given in \cite{Wang2019Transport}, rare fluctuations are
\begin{equation}\label{rare01}
P(x,t)\sim \frac{(\tau_0)^\alpha}{at^\alpha}(\alpha(1-\frac{x/a}{t/\langle\tau\rangle})^{-1-\alpha}-(\alpha-1)(1-\frac{x/a}{t/\langle\tau\rangle})^{-\alpha}),
\end{equation}
describing the scaling when $x$ is of the order of the time $t$.
Eq.~\eqref{rare01} gives a prediction for  particles that are near the initial position. In other words, the mentioned slowing-moving particles stem from these longest single waiting times drawn from Eq.~\eqref{23eqs100}. As shown in Fig.~\ref{RareEvents}, our theory is effective for both the positional distribution's central part and far tails, whereas the L{\'e}vy stable law Eq.~\eqref{23eqs400} tends to infinity.

\begin{figure}[htb]
 \centering
 \includegraphics[width=0.5\textwidth]{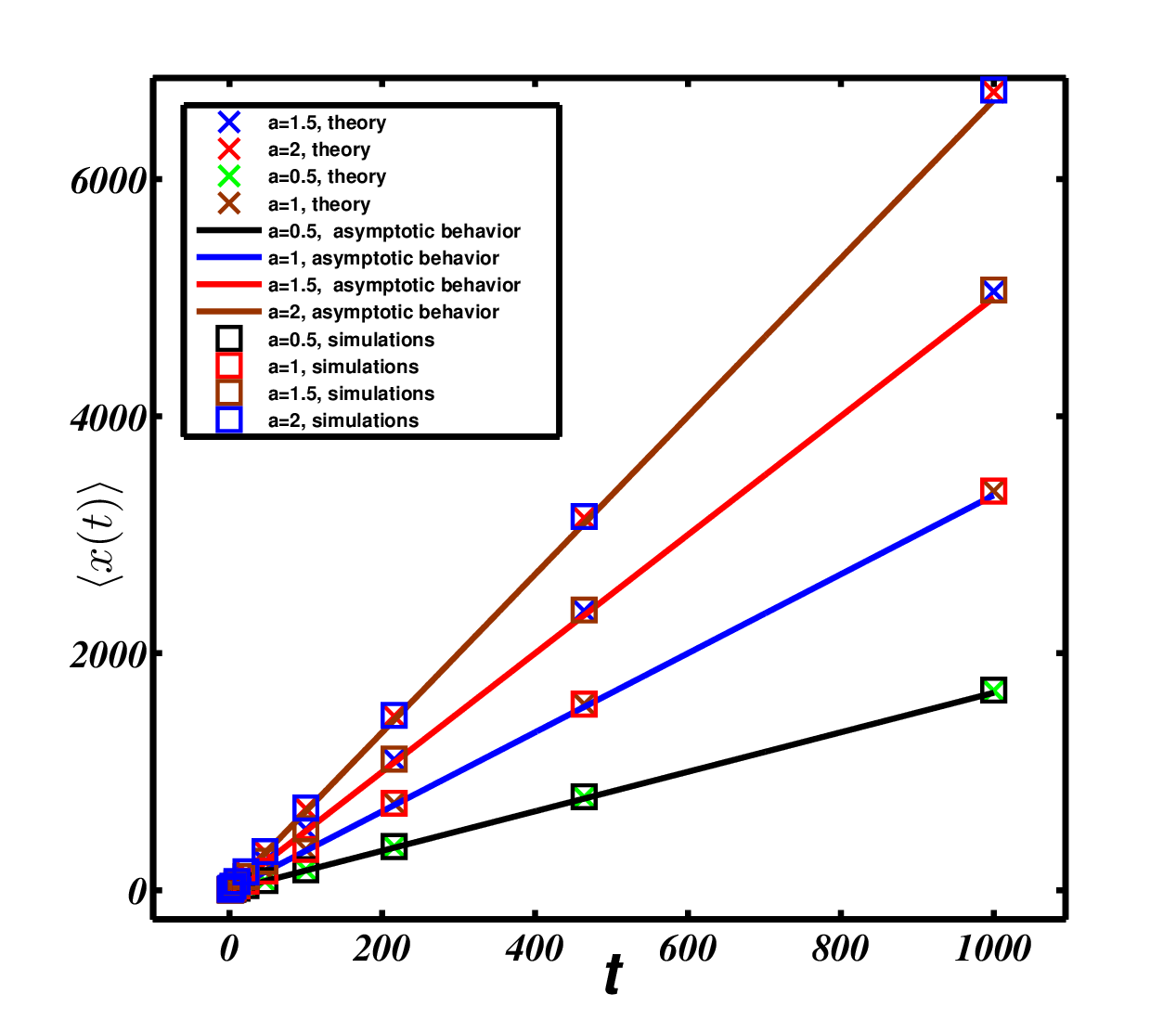}\\
 \caption{Mean of the position under various biases are depicted.
 Lines correspond to the asymptotic prediction given by Eq.~\eqref{mean}. The parameters are $\sigma=1$, $\alpha=3/2$, $\tau_0=0.1$, and $2\times 10^7$ realizations.
}\label{MeanOfXLinear}
\end{figure}

\begin{figure}[htb]
 \centering
 \includegraphics[width=0.5\textwidth]{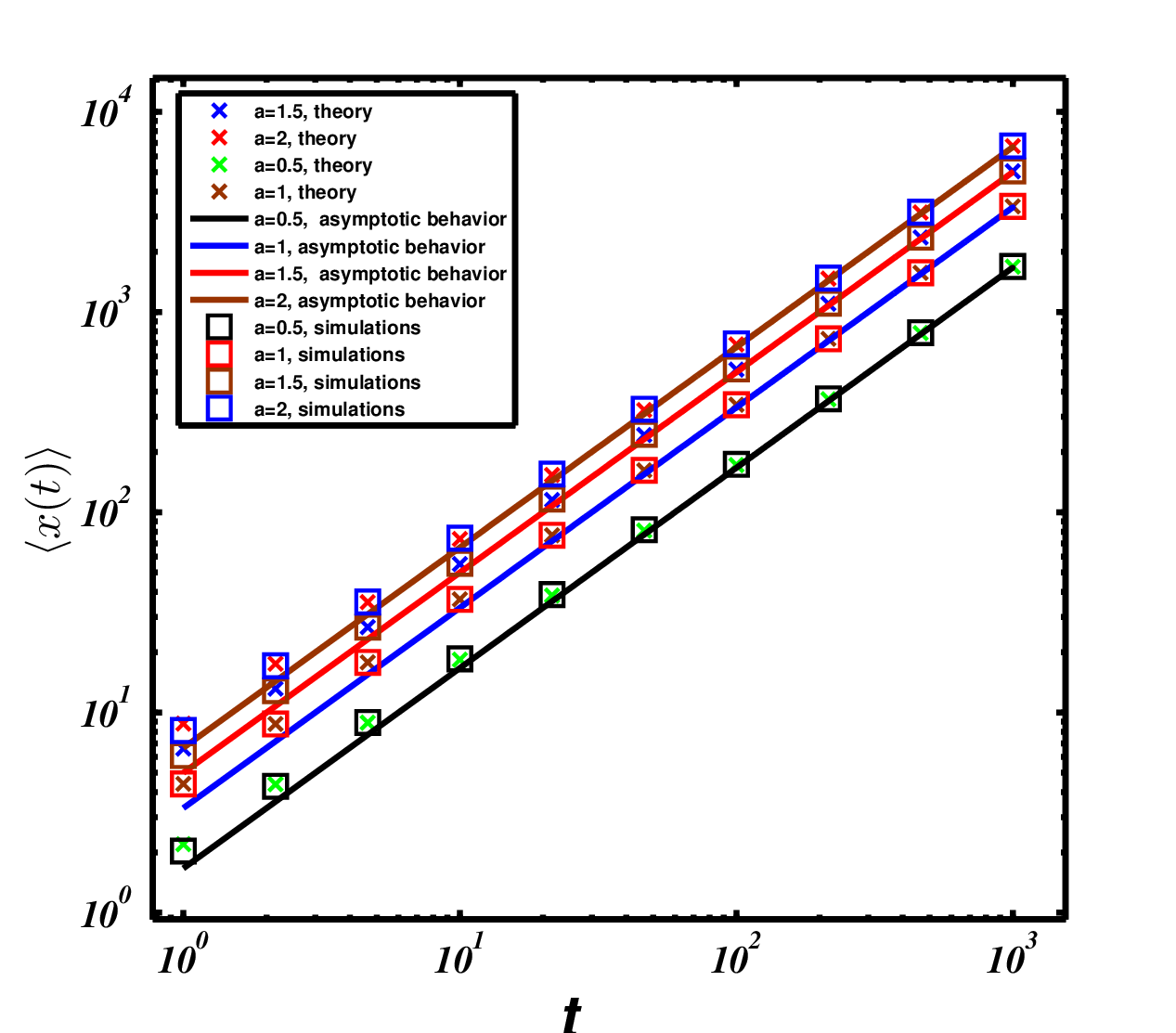}\\
 \caption{Same as Fig.~\ref{MeanOfXLinear} in log-log scale.
}\label{MeanOfXLog}
\end{figure}

\subsection{Moments of the position}
We investigate the mean of the position denoted by $\langle x(t) \rangle$, which refers to the sum of all positions divided by the number of particles. It is a measure that reflects the tendency of the position.  Let us start from the Montroll-Weiss equation, which in Fourier-Laplace space is as follows \cite{Bouchaud1990Anomalous,Metzler2000random}
\begin{equation}\label{23eqs1000}
\widetilde{\widehat{P}}(k,s)=\frac{1-\widehat{\phi}(s)}{s}\frac{1}{1-\widetilde{f}(k)\widehat{\phi}(s)}.
\end{equation}
Using the relation
\begin{equation*}
\langle \hat{x}^q(s)\rangle=i^q\frac{\partial^q \widetilde{\widehat{P}}(k,s) }{\partial k^q}|_{k=0},
\end{equation*}
in the long time limit Eq.~\eqref{23eqs1000} yields
\begin{equation}\label{mean}
\langle x(t)\rangle \sim a\frac{t}{\langle \tau\rangle}.
\end{equation}
Recall that the waiting time has a finite mean $\langle \tau\rangle$, then for a fixed $t$, the mean of the number of renewals is $\langle N(t)\rangle\sim t/\langle\tau\rangle$ \cite{Godreche2001Statistics,Wang2018Renewal}. Thus $\langle x(t)\rangle\sim \langle \triangle x\rangle \langle N(t)\rangle\sim at/\langle \tau\rangle$ is obtained again, where $\langle \triangle x\rangle$ is the mean of the displacements. As shown in Figs.~\ref{MeanOfXLinear} and \ref{MeanOfXLog}, the theoretical prediction grows linearly with the observation time $t$, showing a perfect match with simulations and Eq.~\eqref{mean}.

\begin{figure}[htb]
 \centering
 \includegraphics[width=0.5\textwidth]{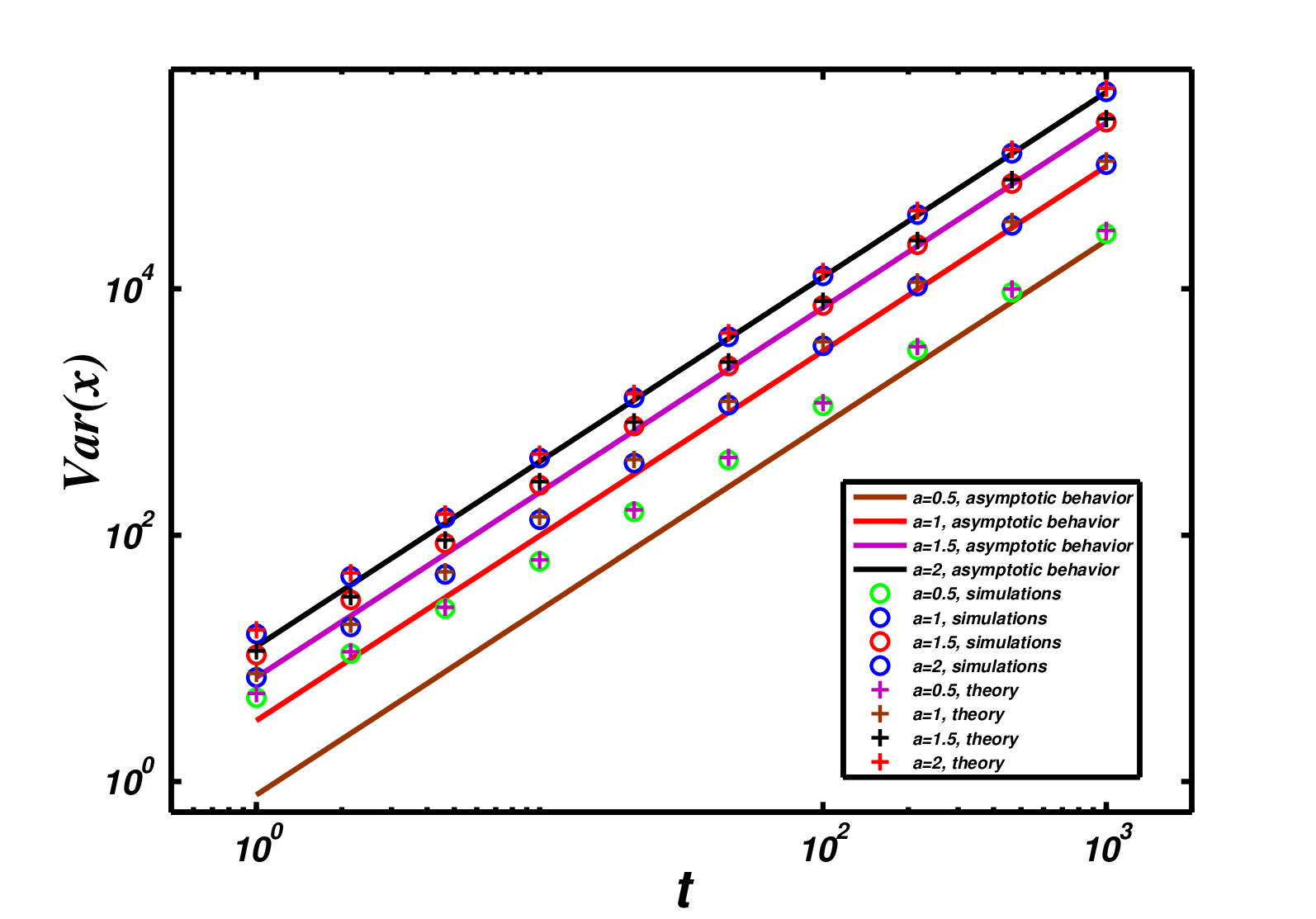}\\
 \caption{Plot of the Var$(x)$. The solid lines describes the asymptotic behavior Eq.~\eqref{23eqs1003}, together with simulations ('$\circ$') and theories ('$+$'). The parameters are the same as in Fig.~\ref{MeanOfXLinear}.
}\label{VarOfXLog}
\end{figure}

\begin{figure}[htb]
 \centering
 \includegraphics[width=0.5\textwidth]{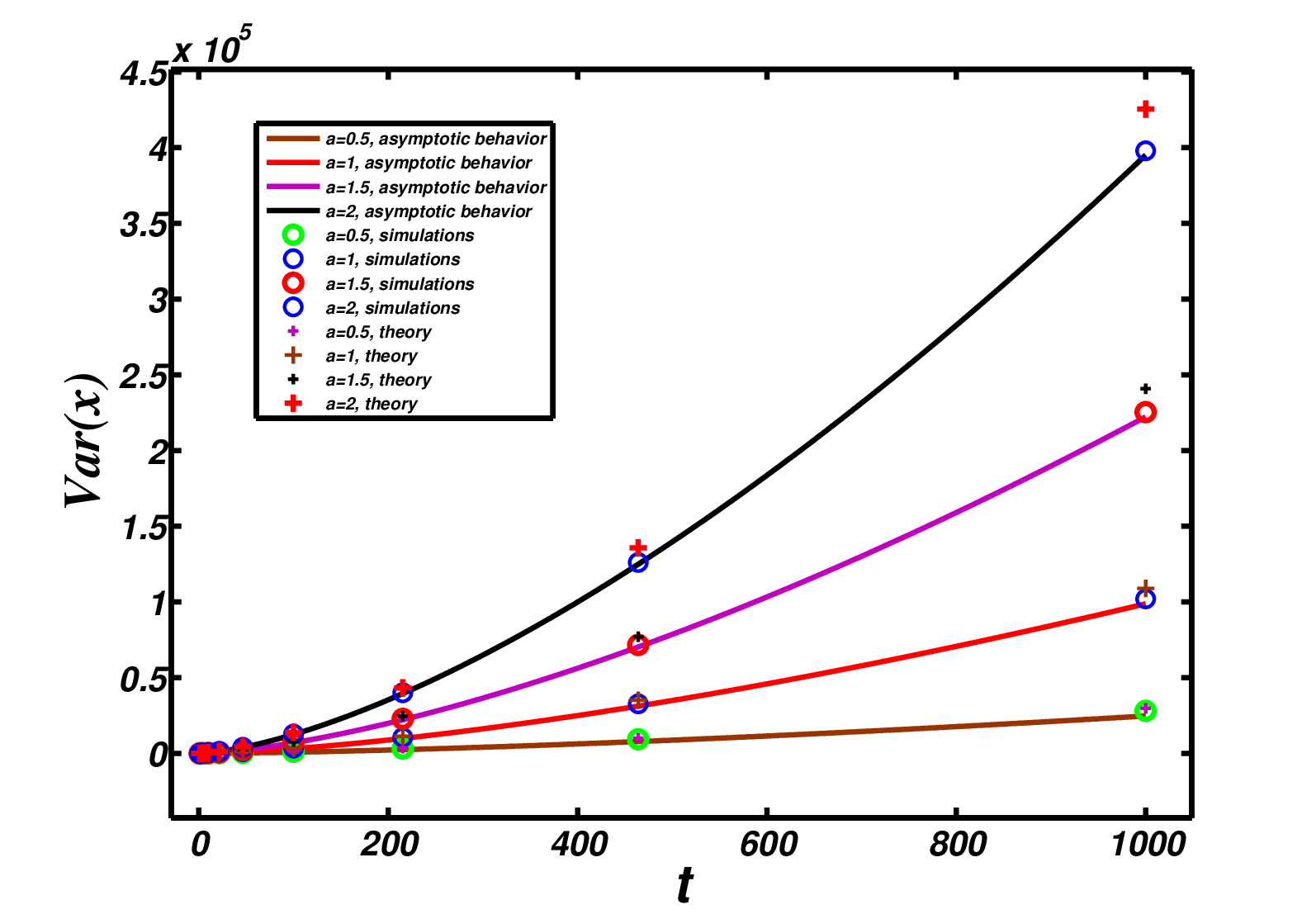}\\
 \caption{Same as in Fig.~\ref{VarOfXLog} in linear-linear scale.
}\label{VarOfXLinear}
\end{figure}

The MSD is the most commonly employed quantifier for diffusion processes. Due to the effect of the bias, the biased CTRW shows superdiffusion \cite{Hou2018Biased}. Fascinatingly, the MSD is controlled by these sluggish particles that significantly lag far behind the average position, as opposed to the swift particles near the average. In that sense, the rare events, governed by the infinite density, are responsible for the enhanced diffusion \cite{Wang2019Transport}.  Mathematically, the variance of the position grows as
\begin{equation}\label{23eqs1003}
\begin{split}
    {\rm Var}(x)&=\langle x^2\rangle-\langle x\rangle^2\\
&\sim \frac{2a^2\tau_0^\alpha}{(2-\alpha)(3-\alpha)\langle\tau\rangle^3}t^{3-\alpha}+\sigma^2\frac{t}{\langle\tau\rangle},
\end{split}
\end{equation}
measuring the time evolution of its spatial extension. In the long time limit, the leading term of Eq.~\eqref{23eqs1003} is $t^{3-\alpha}$, and when the observation time is short, the correction term is of importance. As shown in Fig.~\ref{VarOfXLog}, theoretical results are consistent with simulations. In particular, when the bias is weak, for example, $a=0.5$, our theory gives a perfect prediction if compared with Eq.~\eqref{23eqs1003}.
In linear-linear scale, small deviations exist between  theoretical predictions and simulation results; see Fig.~\ref{VarOfXLinear}. The reason is as follows. When the bias is strong, particles are pushed by the bias and move quickly. In this scenario, the asymptotic L{\'e}vy stable law given by Eq.~\eqref{23eqs400} applies. Consequently, the modified infinite density becomes insignificant, as its probability is very low compared to the weak bias case (see Fig.~\eqref{PxofLog}). Therefore, the algorithm provided in Sec. \ref{Algo} essentially utilizes Eq.~\eqref{23eqs400} with a cutoff at the tail to compute the MSD. Besides, fluctuations of the far tail of the positional distribution are large when the bias are strong. It leads to deviations, but it does not like the results given by Eq.~\eqref{23eqs401}, which predicts an infinite MSD.

%One of the reasons may be that when $\xi > \xi^{*}$, we utilize the information from the asymmetric L{\'e}vy stable distribution instead of the infinite density given in Eq.~\eqref{app102}.
%the asymptotic prediction Eq.~\eqref{23eqs1003} nearly fails, but

\subsection{Breakthrough curves}
Now we further discuss the positional distribution, but from aspects of breakthrough curves, which are essential tools in measuring the contamination in porous materials \cite{Berkowitz2006Modeling,Berkowitz2009Exploring,Alon2017Time,Generalized2022Ricardo,Doerries2022Rate}. We assume that all the particles are injected immediately when we start to observe the process, wherein the positional concentration is measured at a fixed position $x_{b}$. To be more exact, the breakthrough curves, the probability of particles being at $x_b$ at a specific observation time $t$, are measured in terms of the distribution.

\begin{figure}[htb]
 \centering
 \includegraphics[width=0.5\textwidth]{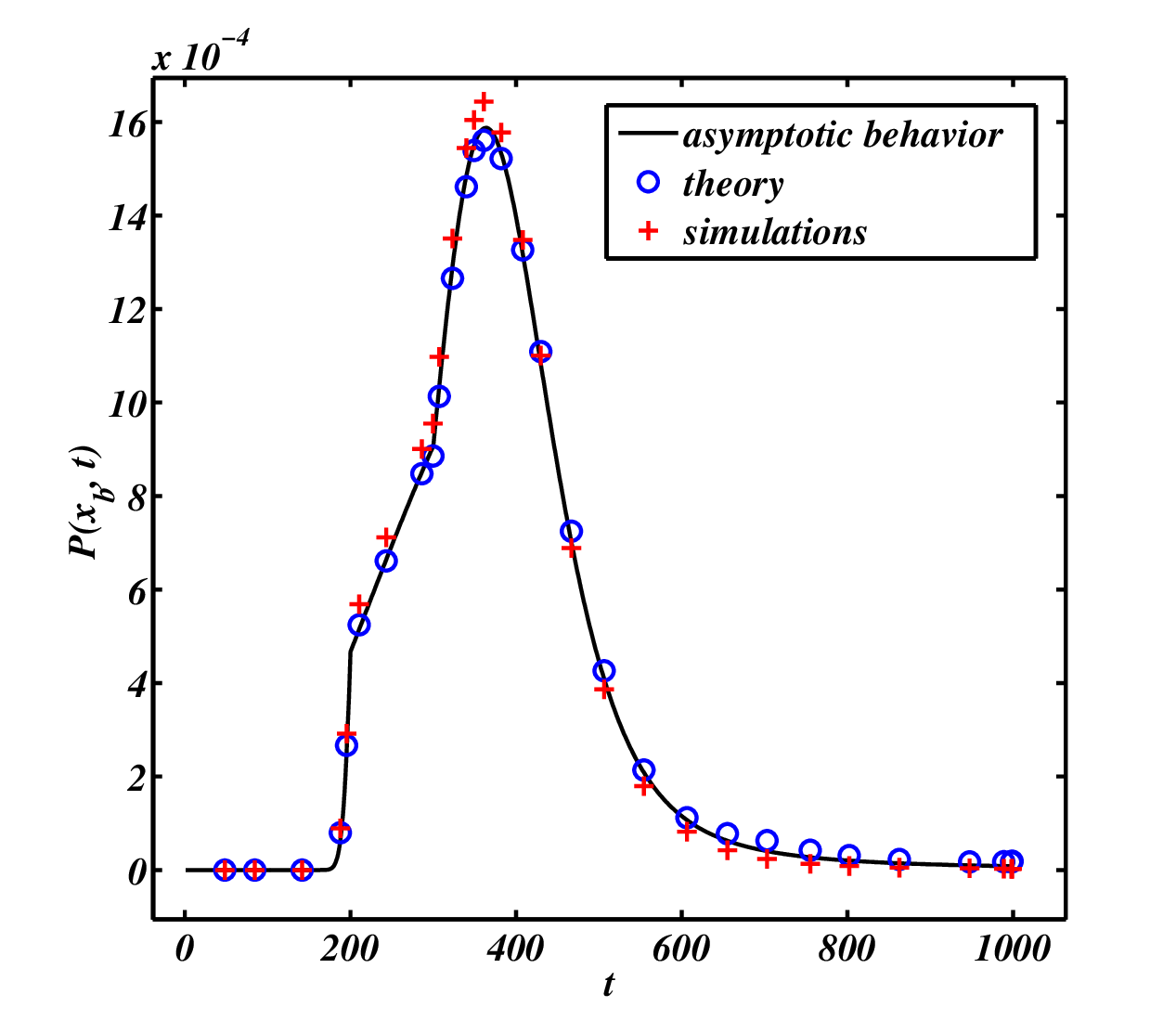}\\
 \caption{Breakthrough curves with time-dependent bias. The theory is based on the algorithm given in Sec. \ref{set2}, asymptotic behavior is given in Eq.~\eqref{aaeqfk305}, and simulations are obtained from $2\times 10^7$ trajectories. Here we choose $t_a=t_b/2=t_c/3=100$, $t=1000$, $\eta=4$, and $\sigma=5$.
}\label{breakthroughCurvesxi15}
\end{figure}

Based on the discussions in \cite{Alon2017Time,Wang2020Fractional}, we delve into the time-dependent bias to validate the program outlined in Sec. \ref{set2}. The process entails four states. In state (i), particles are promptly injected; thereafter, displacements follow a Gaussian distribution with a mean of $a$ and a variance of $\sigma^2$ from time zero until $t_a$. All along the process, the variance of displacements remains the same. Subsequently, in state (ii), the bias of the system is enhanced from time $t_a$ to $t_b$, i.e., the displacements are drawn from Gaussian distribution with the mean $4a\eta/(1+\eta)$, where $\eta>2/3$. State (iii) witnesses a bias of the system represented by $a/(2/3+\eta)$ over the time interval $(t_b,t_c)$, constituting the weakest bias among the four states. Ultimately, the bias reverts to $a$ after $t_c$ until the end of the observation time. The corresponding theoretical prediction was considered using the fractional advection-diffusion-asymmetry equation in Ref.~\cite{Wang2020Fractional}, where the main strategy is that the final position of each state is regarded as the initial position of the next process. In the long time limit, the positional distribution follows \cite{Wang2020Fractional}
\begin{widetext}
\begin{equation}\label{aaeqfk305}
P(x,t)\sim\int_{-\infty}^{\infty}\frac{1}{\sqrt{4\pi c_{m1}}}\exp\left[-\frac{(x-y-c_{m2})^2}{4c_{m1}}\right]\frac{1}{(c_{m3})^{1/\alpha}}\mathcal{L}_{\alpha}\left[\frac{y}{(c_{m3})^{1/\alpha}}\right]{\rm d}y
\end{equation}
with
\begin{equation}\label{aaeqfk302}
c_{m1}=\left\{
  \begin{array}{ll}
   t\frac{\sigma^2}{2\langle\tau\rangle} , & \hbox{$m=1$;} \\
   t\frac{\sigma^2}{2\langle\tau\rangle}, & \hbox{$m=2$;} \\
   t\frac{\sigma^2}{2\langle\tau\rangle}, & \hbox{$m=3$;} \\
   t\frac{\sigma^2}{2\langle\tau\rangle} , & \hbox{$m=4$,}
  \end{array}
\right.
\end{equation}

\begin{equation}\label{aaeqfk303}
c_{m2}=\left\{
  \begin{array}{ll}
   a_1\frac{t}{\langle\tau\rangle}, & \hbox{$m=1$;} \\
   \left[a_1t_1+a_2(t-t_1)\right]\frac{1}{\langle\tau\rangle}, & \hbox{$m=2$;} \\
  \left[a_1t_1+a_2(t_2-t_1)+a_3(t-t_2)\right]\frac{1}{\langle\tau\rangle}, & \hbox{$m=3$;} \\
  \left[a_1t_1+a_2(t_2-t_1)+a_3(t_3-t_2)+a_4(t-t_3)\right]\frac{1}{\langle\tau\rangle} , & \hbox{$m=4$,}
  \end{array}
\right.
\end{equation}
and
\begin{equation}\label{aaeqfk304}
c_{m3}=\left\{
  \begin{array}{ll}
   a_1^\alpha t\frac{1}{\overline{t}}, & \hbox{$m=1$;} \\
   \left[a_1^\alpha t_1+a_2^\alpha(t-t_1)\right]\frac{1}{\overline{t}}, & \hbox{$m=2$;} \\
  \left[a_1^\alpha t+a_2^\alpha(t_2-t_1)+a_3^\alpha(t-t_2)\right] \frac{1}{\overline{t}}, & \hbox{$m=3$;} \\
 \left[a_1^\alpha t_1+a_2^\alpha(t_2-t_1)+a_3^\alpha(t_3-t_2)+a_4^\alpha(t-t_3)\right] \frac{1}{\overline{t}}, & \hbox{$m=4$.}
  \end{array}
\right.
\end{equation}
\end{widetext}
with $m=1,2,3,4$ being the number of the state. This is plotted in  Fig.~ \ref{breakthroughCurvesxi15} for different times $t$. Note that when $0<t<t_1$, the solution Eq.~\eqref{aaeqfk305} reduces to  Eq.~(7) in the main text.

\section{Conclusion}\label{24seccon}
In this manuscript, we introduced a way to generate statistics of the positional distribution that is consistent with the CTRW model for a long observation time $t$. The main idea is that we use the modified L{\'e}vy stable law  $\mathcal{L}^{*}_{\alpha}(\xi)$ to simulate the number of renewals, which yields a perfect prediction for the positional distribution. In some sense, the method avoids the generating of IID random waiting times, which saves a significant amount of time.
Mathematically, we can use the following formula to predict the positional distribution
\begin{equation}\label{18eq503con}
\begin{split}
P(x,t)\sim &\int_{-\frac{t}{\langle\tau\rangle}(\overline{t}/t)^{1/\alpha}}^{\infty}\mathcal{L}^{*}_{\alpha}(\xi)\\
    & \times \frac{\exp\Big(-\frac{(x-a\frac{t}{\langle\tau\rangle}-a\xi(\frac{t}{\overline{t}})^{1/\alpha})^2}{2\sigma^2(t/\langle\tau\rangle+\xi(t/\overline{t})^{1/\alpha})}\Big)}{\sqrt[]{2\sigma^2\pi(\frac{t}{\langle\tau\rangle}+\xi(\frac{t}{\overline{t}})^{1/\alpha})}}{\rm d}\xi,
\end{split}
\end{equation}
with
\begin{equation}
\mathcal{L}^{*}_{\alpha}(\xi)=\left\{
                                \begin{array}{ll}
                                  \mathcal{L}^{*}_{\alpha}(\xi), & \hbox{$\xi\geq \xi^{*}$;} \\
Q_t^{I}(\xi), & \hbox{$\bar{b}<\xi<\xi^{*}$}\\
                                  \bar{b}, & \hbox{else.}
                                \end{array}
                              \right.
\end{equation}
The method is not limited to examples discussed in the manuscript. On the one hand, it is valid for the case of waiting times having an infinite mean, where the number of renewals follows the one-sided L{\'e}vy stable distribution \cite{Godreche2001Statistics}. On the other hand, the displacements can be extended to situations when steps have a finite mean and a finite variance. When discussing the MSD, deviations exist between simulations and theoretical predictions. However, we did not find such deviations for the positional distribution and the mean of the position. This indicates that, for the biased CTRW model, the MSD is sensitive to the details of the positional distribution. Improving the presented work is a task left for the future.

Recently, researchers have been increasingly drawn to the nearly exponential decay observed in stochastic processes \cite{Kege2000Direct,Masolivera200dynamic,Weeks2000Three,Pinaki2007Universal,Wang2009Anomalous,Hapca2009Anomalous,Leptos2009Dynamics,Eisenmann2010Shear,Toyota2011Non,Skaug2013Intermittent,Xue2016Probing,Wang2017Three,Jeanneret2016Entrainment,Chechkin2017Brownian,Cherstvy2019Non,Witzel2019Heterogeneities,Shin2019Anomalous,Singh2020Non,Mejia2020Tracer,Xue2020Diffusion,defaveri2023stretchedexponential,Hu2023Triggering}. Unlike the approach investigated here, the exponential decay is easy to detect at a short time scale, such as, the total observation time is half. Consequently, Eq.~\eqref{18eq503con} fails. How to extend the present approach is left for future work.

\section*{Acknowledgments}
WW expresses gratitude to Eli Barkai for the discussion that inspired this work.
The work is supported by the National Natural Science Foundation of China under Grant No.
12105243 and 12101555, and the Zhejiang Province Natural Science Foundation LQ$22$A$050002$.
%\bibliographystyle{prestyle}
%\bibliography{wenxian}

%\section{Generated random variables}
%\subsection{Generated random variables from Infinite density}

\end{document}